\def\jdc{J_{\rm DC}}
\documentclass{jpsj2}
\usepackage{epsf}
\usepackage{graphicx}
\usepackage{amsfonts}
\usepackage{dcolumn}%
\usepackage{color,ulem}
\usepackage{bm}
\begin{document}

\title{Zero-Field Fiske Resonance Coupled with Spin-waves in Ferromagnetic Josephson Junctions}

\author{Shin-ichi Hikino$^{1,3}$, Michiyasu Mori$^{2,3}$, and Sadamichi Maekawa$^{2,3}$}
\inst{
$^{1}$Computational Condensed Matter Physics Laboratory, RIKEN, Wako, Saitama 351-0198, Japan \\
$^{2}$Advanced Science Research Center, Japan Atomic Energy Agency, Tokai, Ibaraki 319-1195, Japan \\
$^{3}$CREST, Japan Science and Technology Agency (JST), Kawaguchi, Saitama 332-0012, Japan
} 

\date{\today}

\abst{
AC Josephson current density in a Josephson junction with DC bias is spatially modulated by an external magnetic field, and induces an electromagnetic (EM) field inside the junction.
The current-voltage ($I$-$V$) curve exhibits peaks due to the resonance between the EM field and the spatially modulated AC Josephson current density. 
This is called {\it Fiske resonance}. 
Such a spatially modulated Josephson current density can be also induced 
by a non-uniform insulating barrier and the Fiske resonance appears without external magnetic field. This is called zero-field Fiske resonance (ZFFR).
In this paper, we theoretically study the ZFFR coupled with spin-waves in a superconductor/ferromagnetic insulator/superconductor 
junction (ferromagnetic Josephson junction) with a non-uniform 
ferromagnetic insulating barrier. 
The resonant mode coupled with spin-waves can be induced without external magnetic field. 
We find that the $I$-$V$ curve shows resonant 
peaks associated with composite excitations of spin-waves and the EM field in the junction. 
The voltage at the resonance is obtained as a function of the normal modes of EM field. 
The ZFFRs coupled with spin-waves are found as peak structures in the DC Josephson current density as a function of bias voltage. 
}


\maketitle

\section{Introduction}
The DC Josephson effect is characterized by the  
DC current flowing without a voltage-drop between two superconductors separated by a thin insulating barrier~\cite{josephson}. 
When a DC voltage $V$ is applied to the junction, 
the AC Josephson current with frequency $(2e/\hbar)V$ flows in the junction driven by  
the difference of phases in two superconducting order parameters, i.e., {\it Josephson-phase} $\theta$.
If both the DC voltage and a magnetic field are applied to the junction, 
whose width $L$ is smaller than the Josephson penetration depth $\lambda_{\rm J}$, 
the AC Josephson current density is spatially modulated and generates the electromagnetic (EM) field inside the junction.
In this case, the current-voltage ($I$-$V$) curve exhibits peaks due to the resonance 
between the AC Josephson current density and the EM field. 
This is called {\it Fiske resonance}~\cite{fiske, eck, coon, kulik, barone}. 

Josephson junctions composed of ferromagnetic metal (FM) and superconductors (Ss) are extensively studied for the last decade. 
The S/FM/S junctions exhibit fascinating phenomena which are not observed in the conventional Josephson junctions 
\cite{golubov_sfs1, buzdin_sfs1, ryazanov_sfs1, kontos_sfs1, robinson}. 
The interaction between Cooper pairs and spin waves in the FM is of importance 
in the transport properties in the S/FM/S and S/I/FM/S junctions. 
The dynamics of $\theta$ coupled with spin-waves in the FM 
has been investigated theoretically 
\cite{nussinov_sfs1,stakahashi_sfs1,bell_sfs1,houzet_sfs1,hikino_sfs1,konschelle_sfs1,yokoyama_sfn1,volkov_sfs1,mai-sfs} and experimentally~\cite{petkovic_sfs1}. 
However, the Fiske resonance coupled with spin-waves is not yet observed experimentally. 

Another type of Josephson junction with ferromagnetic insulator (FI) instead of the FM is also examined. 
It is reported that the dissipation effect in the S/FI/S junction is smaller than that the S/FM/S junction~\cite{kawabata1,kawabata2,note}. 
The damping of spin-waves is also very small in the FI compared to the case of the FM \cite{hillebrands,stancil,kajiwara}. 
Therefore, the coupling between $\theta$ and spin-waves can be observed more clearly in the S/FI/S junction. 
In fact, in the S/FI/S junction, it is expected that the Fiske resonance has clear multiple structures associated with 
spin-wave excitation\cite{hikino-sfis}. 

Here, we note that the Fiske resonance in the conventional Josephson junction 
is also induced by the non-uniform insulating barrier in the junction, since AC Josephson current density driven by a DC voltage is spatially modulated 
and then the EM field is generated inside the junction. 
In this case, the Fiske resonance occurs without external magnetic field. 
It is called zero-field Fiske resonance (ZFFR), which originates from the resonance between the EM field and the spatially modulated AC Josephson current density due to the non-uniform insulating barrier. 
This phenomenon has been widely studied experimentally and theoretically in the Josephson junction~\cite{barone,russo,wang,camerlingo,nappi}. 

In this paper, we theoretically study the ZFFR coupled with spin-waves in an S/FI/S junction with a non-uniform FI. 
The merit of such a non-uniform geometry of junction is that the spatially modulated AC Josephson current density can be induced with no external magnetic field 
and thus the Fiske resonance occurs without external magnetic field. 
By solving the equation of motion of $\theta$ coupled with spin-waves, 
it will be found that the $I$-$V$ curve shows resonant peaks. 
The voltage at the resonances is obtained as a function of the normal modes of EM field, 
which indicates composite excitations of the 
EM field and spin-waves in the S/FI/S junction. 
Dependence of those resonances on distributions of the Josephson critical current density is presented. 

The rest of this paper is organized as follows. 
In Sec. II, we formulate 
the Josephson current in a Josephson junction with a non-uniform ferromagnetic insulator.
In Sec. III, 
the DC component of Josephson current density at the ZFFR with spin-waves is calculated analytically and numerically. 
Summary is given in Sec. IV. 

\begin{figure}[t]
\begin{center}
\vspace{10mm}
\includegraphics[width=6cm]{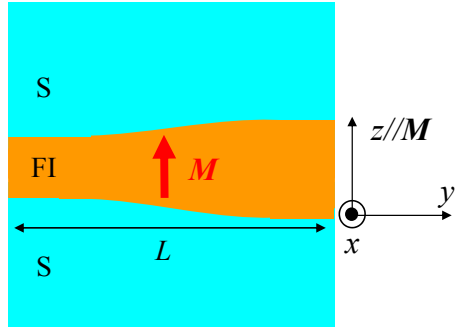}
\caption{ (Color online) Schematic of a superconductor/ferromagnetic insulator/superconductor (S/FI/S) junction 
with the magnetization $\bm M$ in the FI. 
The non-uniform geometry of the junction with the width $L$ is schematically illustrated. 
}
\label{gm-fjj}
\end{center}
\end{figure}

\section{FORMULATION of JOSEPHSON CURRENT in FERROMAGNETIC JOSEPHSON JUNCTION with MAGNETIC INSULATOR}
The system considered is a Josephson junction with a FI sandwiched 
by two superconductors with $s$-wave symmetry as shown in Fig.$~$\ref{gm-fjj}. 
The geometry of the junction is assumed to be a non-uniform junction to impose non-uniform Josephson current density 
without external magnetic field~\cite{note1}.
The magnetization in the FI is parallel to the $z$-direction. 
Here, we assume a simple model of non-uniform Josephson current density given by, 
\begin{eqnarray}
J \left( {y,t} \right) &=&
 J_{\rm{c}} \left( y \right)\sin \left[ {\omega _{\rm J} t + \theta \left( {y,t} \right)} \right], \\
J_{\rm{c}} \left( y \right) &=& 
J_{
\rm{0}.
} 
P
 \left( y \right)
\left[ \left( {1 - \zeta } \right)\frac{{\cosh \left[ {\kappa \left( {1 - 2y/L} \right)} \right]}}{{\cosh \left( \kappa  \right)}} 
\right. \nonumber \\
&+& 
\left.
\zeta \frac{{\sinh \left[ {\kappa  \left( {1 - 2y/L} \right)} \right]}}{{\sinh \left( \kappa   \right)}} \right] 
\label{jcy}, \\
P \left( y \right) &=& \left\{ \begin{array}{l}
 1 
\,\,\,{\rm for} \,\,\, 0 \le y \le L, \\ 
 0 
\,\,\,{\rm for} \,\,\, y < 0\,\, {\rm or}\,\, L < y, \\ 
 \end{array} \right.
\end{eqnarray}
where $J_{\rm c}(y)$ and $\omega_{\rm J}=(2e/\hbar)V$ are the Josephson critical current density and Josephson frequency 
with bias voltage $V$, respectively~\cite{note3}. 
$J_{0}$ is the Josephson critical current density for the uniform geometry of junction, i.e., $\zeta=\kappa=0$. 
The electromagnetic dynamics induces $\theta(y,t)$ depending on space and time.
The distribution of $J_{\rm c}(y)$ is determined by two parameters $\kappa$ and $\zeta$, where $0\leq \zeta \leq 1$ is imposed. 
In the S/FI/S junction, 
spin-waves can be excited by the EM field inside the FI due to the AC Josephson current. 
In this situation, the equation of motion for $\theta(y,t)$ coupled with spin waves is described by~\cite{hikino-sfis}, 
\begin{eqnarray}
\frac{{\partial ^2 \theta \left( {y,t} \right)}}{{\partial y^2 }} &=& 
\frac{1}{{c^2 }}
\left[ {
{\frac{{\partial ^2 \theta \left( {y,t} \right)}}{{\partial t^2 }}}
+ \frac{1}{\mu_{0}} 
\int_{-\infty }^{\infty } dy'dt' 
\chi (y-y',t-t') \frac{{\partial ^2 \theta \left( {y',t'} \right)}}{{\partial t'^2 }} 
} \right. \nonumber \\
&+& 
\left. {
\Gamma \frac{\partial \theta(y,t)}{\partial t}
+ \Gamma \frac{1}{\mu_{0}}
\int_{-\infty }^{\infty } dy'dt' 
\chi (y-y',t-t') \frac{{\partial \theta \left( {y',t'} \right)}}{{\partial t' }} 
 } \right] \nonumber \\
&+&
\frac{1}{\lambda_{\rm J}^{2}\left<J_{\rm c}(y)\right>} 
J(y,t)
+ \frac{1}{\lambda_{\rm J}^{2}\left<J_{\rm c}(y)\right>} \frac{1}{\mu_{0}}
\int_{-\infty }^{\infty } dy'dt' 
\chi (y-y',t-t') 
J(y',t'),
\label{sgeq-sw} \\
\left<J_{\rm c}(y)\right> &=& \frac{1}{L} \int_{0}^{L} dy J_{\rm c}(y).
\label{av-jc}
\end{eqnarray}
The effective velocity of light in the FI $c$ is given by ${c} = \sqrt {d/[(d+2\lambda_{\rm L})\epsilon \mu_{0}]}$ 
, Josephson penetration depth $\lambda_{\rm J} = \sqrt{\hbar/[2e\mu_{0}(d + 2\lambda_{\rm L})J_{\rm 0}]}$, dielectric constant $\epsilon$ and permeability $\mu_0$. 
The London penetration depth is denoted by $\lambda_{\rm L}$ and 
$\Gamma \equiv (\epsilon R)^{-1}$ means the damping factor caused by quasi-particle resistivity $R$ in the FI. 
The magnetic susceptibility of the FI in the linearized Landau-Lifshitz-Gilbert equation is given by~\cite{hillebrands},
\begin{eqnarray}
\chi \left(q, \omega_{\rm J} \right) = 
		\gamma M_{z}
		\frac{\Omega_{\rm S} + i \alpha \omega_{\rm J}}
		{\Omega_{\rm S}^{2} - (1+\alpha^{2})\omega_{\rm J}^{2} + i2\alpha \Omega_{\rm S} \omega_{\rm J}}
\label{chi-f},
\end{eqnarray}
where $M_{z}$, $\alpha$, and $\gamma$ are the $z$-component of the magnetization, Gilbert damping factor, and the gyromagnetic ratio, respectively. 
Magnetic susceptibility and spin-wave energy $\hbar \Omega_{\rm S}$ in a magnetic material are generally modified by geometry and thickness. 
On the other hand, Eq.~(\ref{chi-f}) is obtained by assuming a uniform FI.   
This is justified, because the magnetic susceptibility and $\hbar \Omega_{\rm S}$ are insensitive to the thickness of FI, 
provided that the conformation of the ferromagnetic materials changes on a scale of nanometers~\cite{sun}. 
Therefore, we adopt Eq.~(\ref{chi-f}) and $\hbar \Omega_{\rm S}$ obtained in the uniform FI~\cite{note2} 
for the non-uniform FI as an approximation, since we consider 
the thickness change of FI to be in a range of a few nanometers. 
In the FI, the dispersion relation of spin-waves with the frequency $\Omega_{\rm S}$ is given by 
\begin{eqnarray}
\Omega_{\rm S} &=& \Omega_{\rm B} + \frac{\eta}{\hbar} q^{2}
\label{omega-sw}, 
\end{eqnarray}
where $\Omega_{\rm B} = \gamma (H_{\rm K} - M_{z}/\mu_{0})$. 
The anisotropic field and the stiffness of spin-waves in the FI are denoted by $H_{\rm K}$ and $\eta$, respectively. 
The spin-wave having a finite wave number $q$ is neglected 
in the Fiske resonance because of the following reason: 
In Eq.$~$(\ref{omega-sw}), the first term $ \Omega_{\rm B}$ is caused by the anisotropic and 
demagnetizing fields, and the wave number $q$ is given by $n \pi/L$. 
In a conventional FI, $\hbar \Omega_{\rm B}$ is about tens of $\mu$eV~\cite{hillebrands}. 
On the other hand, $\eta q^{2}$ is of the order peV due to the small stiffness of spin-waves~\cite{pajda-prb} 
when $L$ is a few mm. 
Below, we only consider $q=0$ mode for spin-waves with the constant frequency $\Omega_{\rm B}$. 

\section{DC Josephson current density with ZFFR and numerical results} 
In order to obtain the solution of Eq~(\ref{sgeq-sw}), 
we expand $\theta(y,t)$ in terms of the normal modes of the EM field generated by the AC 
Josephson current as follows, 
\begin{equation}
\theta(y,t) = {\rm Im}
\left[
\sum_{n=0}^{\infty}
g_{n} e^{i \omega_{\rm J} t} \cos\left( k_{n} y \right)
\right]
\label{sol-theta1}, 
\end{equation}
where $g_{n}$ is a complex number and $k_{n}=n \pi /L$. 
This equation of $\theta(y,t)$ satisfies 
$[\partial \theta/\partial y]_{y=0} = [\partial \theta/\partial y]_{y=L} = 0$, 
which is Kulik's boundary condition~\cite{kulik, barone}. 
We consider $\theta(y,t)$ to be a small perturbation and 
solve Eq.$~$(\ref{sol-theta1}) by taking 
$J(y,t)$ to be $J_{\rm c}(y)\sin(\omega_{\rm J}t)$. 
Substituting Eq.$~$(\ref{sol-theta1}) into Eq.$~$(\ref{sgeq-sw}), 
$g_{n}$ is determined as,
\begin{eqnarray}
	g_{n} &=&	
		-\frac{c^{2} J_{\rm 0}}{\lambda_{\rm J}^{2} \left<J_{\rm c}(y) \right>} \nonumber \\
		&\times& \frac{ 1 + \chi(-\omega_{\rm J})/\mu_{0} }{\omega_{\rm J}^{2} [1+\chi(-\omega_{\rm J})/\mu_{0} ] - \omega_{n}^{2} 
		+i \Gamma \omega_{\rm J}[1+\chi(-\omega_{\rm J})/\mu_{0}]} \nonumber \\
		&\times&
		\left[
		(1-\zeta)B_{n} +\zeta C_{n} 
		\right]
\label{gn}, \\
	B_{n} &=&
		\frac{2 \cos(n\pi/2)}{\cosh(\kappa)}
		\int_{0}^{1} dy
		\cos(n\pi y/2) \cosh(\kappa y), \nonumber \\
	C_{n} &=&
		\frac{2 \sin(n\pi/2)}{\sinh(\kappa)}
		\int_{0}^{1} dy
		\sin(n\pi y/2) \sinh(\kappa y), \nonumber
\end{eqnarray}
where $\omega_{n} = (c \pi/L)n$. 

Next, we calculate the DC Josephson current density $J_{\rm DC}$ coupled with spin waves as a function of $V$. 
The function, $\sin(\omega_{\rm J}t + \theta(y,t))$, is expanded in terms of $\theta(y,t)$ 
and $J_{\rm DC}$ is given by
\begin{equation}
J_{\rm DC} \approx  \mathop {\lim }\limits_{T \to \infty } 
		\frac{1}{T} \int_{0}^{T} dt
		\frac{1}{L} \int_{0}^{L} dy
		J_{\rm c}(y) \cos (\omega_{\rm J} t) \theta(y,t)
\label{def-dc-jc}. 
\end{equation}
Introducing Eqs.~(\ref{sol-theta1}) and (\ref{gn}) into Eq.~(\ref{def-dc-jc}), 
the analytic formula of $J_{\rm DC}$ without external magnetic field is obtained as, 
%
\begin{eqnarray}
	J_{\rm DC} 
		&\approx& 
		\sum_{n=0}^{\infty } 
		\frac{c^{2} \kappa  J_{\rm 0} }{\lambda_{\rm J}^{2} (1-\zeta)\tanh(\kappa )}
		\Psi_{n}(\omega_{\rm J})
		\left[
		(1-\zeta)
		\frac{\kappa   \cos^{2}\left( \frac{n\pi}{2} \right) \tanh(\kappa )}{\kappa^{2} + (n\pi/2)^{2} } 
		+ \zeta  \frac{\kappa  \sin^{2}\left( \frac{n\pi}{2}\right)\tanh^{-1}(\kappa )}{\kappa^{2} + (n\pi/2)^{2} }
		\right]^{2}
\label{jdc},\\
	\Psi_{n}(\omega_{\rm J}) &\equiv& 
	\frac{ \Gamma \omega_{\rm J}[1+2\chi_{1}(\omega_{\rm J})/\mu_{0}]+\omega_{n}^{2} \chi_{2}(\omega_{\rm J})/\mu_{0} 
	+\Gamma \omega_{\rm J} [\chi_{1}^{2}(\omega_{\rm J}) + \chi_{2}^{2}(\omega_{\rm J})]/\mu_{0}^{2}}
	{ \left[
		\omega_{\rm J}^{2}[1+\chi_{1}(\omega_{\rm J})/\mu_{0}]-[\omega_{n}^{2}+\Gamma \omega_{\rm J} 
	\chi_{2}(\omega_{\rm J})/\mu_{0}]
	  \right]^{2} 
	 +\left[ \Gamma \omega_{\rm J}[1+\chi_{1}(\omega_{\rm J})/\mu_{0}] + \omega_{\rm J}^{2}\chi_{2}(\omega_{\rm J})/\mu_{0}
	 \right]^{2} }, 
\label{psi-n}
\end{eqnarray}
where $\chi_{1}(\omega_{\rm J}) = {\rm Re}[\chi(\omega_{\rm J})]$, $\chi_{2}(\omega_{\rm J}) = {\rm Im}[\chi(\omega_{\rm J})]$. 
Equation~(\ref{jdc}) clearly demonstrates that zero-field resonant modes depend on parameters $\kappa$ and $\zeta$ 
which determine the distribution of the Josephson critical current density flowing through the FI. 
Hence, one can easily find that three cases are possible for the zero-field resonance. 
When $\zeta=0$ ($\zeta=1$), the zero-field resonance only appears at even (odd) numbers of $n$.  
On the other hand, when $\zeta\neq 0, 1$, the zero-field resonance appears at all integers $n$. 

Next, we derive a condition for the ZFFR in the present system by analyzing Eq.~(\ref{psi-n}).
When the denominator of $\Psi_{n}(\omega_{\rm J})$ is minimum with respect to $\omega_{\rm J}$, 
$\Psi_{n}(\omega_{\rm J})$ takes a maximum, so that the DC Josephson current exhibits the resonant behavior. 
The DC voltage, at which the resonance occurs, is determined by neglecting the damping term of Eq.~(\ref{psi-n}) as $\alpha=\Gamma =0$. 
Setting the denominator of $\Psi_{n}(\omega_{\rm J})$ to be zero, the voltage is given by 
\begin{eqnarray}
V^{\pm } = \frac{ \hbar }{ 2e } 
		\sqrt{ \frac{ 1 }{ 2 }
		\left[
		\omega_{n}^{2} + \Omega_{\rm S}^{2} + \frac{\gamma M_{z} \Omega_{\rm S}}{\mu_{0}} 
		\pm 
		\sqrt{ \left( \omega_{n}^{2} + \Omega_{\rm S}^{2} +
		\frac{\gamma M_{z} \Omega_{\rm S}}{\mu_{0}} \right)^{2} 
		-4\omega_{n}^{2} \Omega_{\rm S}^{2} } 
		\right]
		}
\label{v-n}. 
\end{eqnarray}
We have two DC voltages, $V^{+}$ and $V^{-}$, at which the ZFFR occurs for each $n$. 
The integer $n$ is determined by the mode of the EM field in the junction. 
Eq.~(\ref{v-n}) clearly shows that there are two dispersion relations, which 
result from the coupling between the EM field and spin-waves in the FI.
Note that the amplitude of ZFFR strongly depends on $\kappa$ and $\zeta$ as we will see in the next section. 

\begin{figure}[!t]
\begin{center}
\vspace{23mm}
\includegraphics[width=7.5cm]{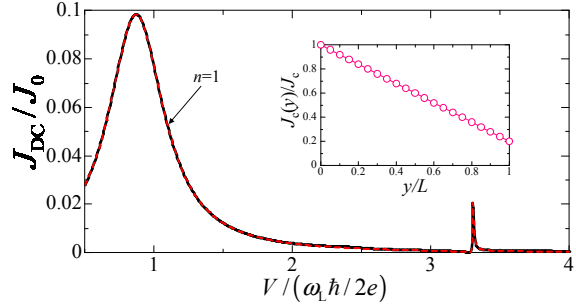}
\caption{ (Color online) DC Josephson current density ($J_{\rm DC}$) as a function of DC voltage ($V$) in 
the S/FI/S junction. 
The black solid line is the total DC Josephson current density. 
The dashed line (Red) is the DC Josephson current density in $n=1$,  
where $n$ is the mode number of EM field (see Eq.~(\ref{jdc})). 
The inset is the distribution of Josephson critical current density as a function of $y$. 
}
\label{fiske1}
\end{center}
\end{figure}

At last, we numerically evaluate Eq.~(\ref{jdc}).  
Parameters are set to be $M_{z}=0.1$ T, $\alpha=1 \times 10^{-4}$~\cite{kajiwara}, $\Omega_{\rm B}/\omega_{\rm L}=3$, 
$\Gamma/\omega_{\rm L}=0.5$~\cite{lisitskiy}, 
$\gamma=2.2\times10^{5}$ m/A$\cdot $s\cite{chikazumi}, and $\omega_{\rm L}\equiv c\pi/L=30$ GHz.  
Instead of plotting an $I$-$V$ curve, $\jdc$ at the resonances will be shown as a function of the voltage $V$ below. 
The amplitude of $\jdc$ is associated with a height of resonant peak 
or a jump in the $I$-$V$ curve (for instance, see Ref\cite{barone}). 
In the following numerical calculations, we exclude the contribution of $n=0$ in Eq.(\ref{jdc}), 
since we discuss about the resonance between the spatially modulated AC Josephson current and standing wave of EM field. 

Figure~\ref{fiske1} shows $\jdc$ induced by ZFFR as a function of $V$~\cite{dis2} for $\kappa=0$ and $\zeta=0.4$. 
With these parameters, $J_{\rm c}(y)$ is linearly distributed in the junction (see the inset of Fig.~\ref{fiske1}). 
The black (solid) and red (dashed) lines are $J_{\rm DC}$ and 
the component with $n=1$ in Eq.~(11), respectively. 
This result clearly demonstrates that the Fiske resonance occurs without external magnetic field, i.e., ZFFR. 
The additional resonance peak around $V/(\omega_{\rm L} \hbar/2e) \approx 3.3$ arises from the presence of spin-wave excitation in the FI. 
This resonance comes from the inhomogeneity of Josephson critical current density induced by the non-uniform geometry of junction. 
Moreover, in the present case, Eq.~(\ref{jdc}) becomes 
\begin{equation}
	J_{\rm DC} =
		\frac{c^{2} J_{\rm 0}}{\lambda_{\rm J}^{2}(1-\zeta)}
		\sum_{n=0}^{\infty }
		\Psi_{n}(\omega_{\rm J})
		\zeta^{2} \left[
		\frac{\sin\left(n\pi/2\right)}{ (n\pi/2) }
		\right]^{4} 
\label{jdc1}. 
\end{equation}
In Eq.~(\ref{jdc1}), it is found that the ZFFR only occurs at odd number of $n$. 
Since resonant peaks of ZFFR with $n>1$ are much smaller than that with $n=1$, 
main contribution to the ZFFR as depicted in Fig.~\ref{fiske1} is the mode of $n=1$. 

\begin{figure}[!t]
\begin{center}
\vspace{23mm}
\includegraphics[width=7.5cm]{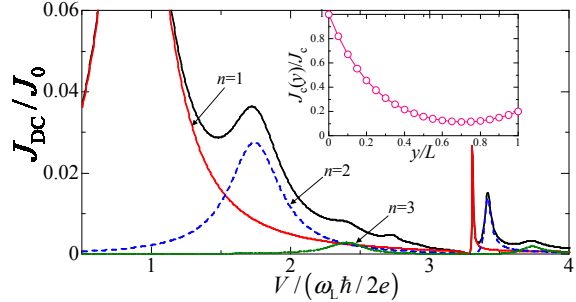}
\caption{ (Color online) DC Josephson current density ($J_{\rm DC}$) as a function of DC voltage ($V$) in 
the S/FI/S junction. 
The black solid line is $J_{\rm DC}$. 
Red (solid), blue (dashed), and green (chain) lines are the DC Josephson current densities in each $n$, 
where $n$ is mode number of EM field (see Eq.~(\ref{jdc})). 
Inset is the distribution of Josephson critical current density as a function of $y$.  
}
\label{fiske2}
\end{center}
\end{figure}

Figure~\ref{fiske2} is the case for $\kappa=2$ and $\zeta=0.4$. 
The black (solid) line is $J_{\rm DC}$. 
Red (solid), blue (dashed), and green (chain) lines are each component  with $n$ in Eq.~(11). 
It is found that ZFFR peaks of $J_{\rm DC}$ clearly appear at $n\geq 1$ in Fig.~\ref{fiske2} in contrast to Fig.~\ref{fiske1}. 
The reason is simply due to the non-linear Josephson critical current density to contain both symmetric and antisymmetric 
components with respect to $y$. 
The present non-linear distribution of the Josephson critical current density will be more realistic. 
Therefore, we can expect that multiple resonant peaks such as Fig.~\ref{fiske2} is practically observed without external magnetic field.

\vspace{5mm}
\section{Summary and Discussion}
We have theoretically studied the zero-field Fiske resonance (ZFFR) in the S/FI/S junction 
with several patterns of spatial variation in the 
Josephson critical current density, which is induced by a non-uniform ferromagnetic insulating barrier. 
Such a non-uniform AC Josephson current density can excite the EM field inside the FI without external magnetic field. 
It is found that the current-voltage ($I$-$V$) curve shows two resonant 
peaks without external magnetic field in the present system, i.e., the ZFFR coupled with spin-waves occurs. 
Voltage at the resonances is obtained as a function of the normal modes of EM field, 
which indicates composite excitations of the EM field and spin-waves in the S/FI/S junction. 

The present study will provide a platform to study the dynamics of Josephson phase and the magnetic excitation. 
Furthermore, in the non-uniform S/FI/S junction, several applications such as spin-current emitter by utilizing spin-wave excitation in the FI
~\cite{maekawa-book} may be also possible in analogy with the emission of coherent THz radiation in the high-T$_c$ cuprate~\cite{Ozyu,Kado,Kosh}, 
although Josephson junctions based on the high-T$_c$ cuprate are usually laminated structures differently from the single Josephson junction discussed here. 
In fact, the inhomogeneity of the junction was one of essential factors to realize the emission 
without external magnetic field~\cite{Ozyu,Kado,Kosh}. However, 
novel devices using the S/FI/S junction are beyond the scope of the present paper and will be studied elsewhere.

\acknowledgements 
This work is supported by Grant-in-Aid for Research Activity Start-up (No. 25887053) from the Japan Society
for the Promotion of Science and Grant-in-Aid for Scientific Research 
from MEXT (Grant No.24540387, No.24360036, No.23340093, and No.25287094), Center for Computational Science and e-Systems of JAEA, and the inter-university cooperative research program of IMR, Tohoku University.



\begin{thebibliography}{}
\bibitem{note}
Both the S/FM/S and the S/FI/S junctions are called ferromagnetic Josephson junction, that means the latter one in this paper.
\bibitem{note1}
The non-uniform geometry of the junction is the essential point, since 
such a geometry introduces the non-uniform current distribution. 
In this paper, the Josephson critical current density 
have a linear current distribution with $y$ in this geometry when $\kappa=0$ and $\zeta \neq 0$.
\bibitem{note3}
Note that we use the voltage bias model for simplicity, since we only focus on the DC component of Josephson current\cite{barone}.
\bibitem{note2}
The distribution of Josephson current density with $y$ (see Eq.~(\ref{jcy})) can be obtained by changing 
the thickness of insulating barrier along $y$ direction, 
since the Josephson current density exponentially decreases with increasing the thickness of insulating barrier.
For instance, see Ref~\cite{barone}.
\bibitem{dis2}
For $L=1$ mm and $c=10^{7}$ $\rm m/\rm s$, the magnitude of DC voltage ($V$) is about tens of $\mu V$.
\end{thebibliography}
\end{document}